\documentclass[twocolumn,superscriptaddress,showpacs,pra]{revtex4}
\usepackage{color}
\usepackage{delarray}
\usepackage{amsmath, amssymb}
\usepackage{bm}
\usepackage{float}
\usepackage{graphicx}

\begin{document}
\title{Rogue wavefunctions due to noisy quantum tunneling potentials}

\author{Cihan Ahmet Bay\i nd\i r}
\email{cihan.bayindir@isikun.edu.tr}
\affiliation{Engineering Faculty, I\c s\i k University,  \.{I}stanbul, Turkey}

\begin{abstract}

In this paper we study the effects of white-noised potentials on nonlinear quantum tunneling. We use a split-step scheme to numerically solve the nonlinear Schr\"{o}dinger equation (NLSE) with a tunneling potential. We consider three different types of potentials, namely; the single rectangular barrier, double rectangular barrier and triangular barrier. For all these three cases we show that white-noise given to potentials do not trigger modulation instability for tunneling of the sech type soliton solutions of the NLSE. However white-noised potentials trigger modulation instability for tunneling of the sinusoidal wavefunctions, thus such a wavefield turns into a chaotic one with many apparent peaks. We argue that peaks of such a field may be in the form of rational rogue wave solutions of the NLSE. Our results can be used to examine the effects of noise on quantum tunneling. Since a rogue wavefunction means a higher probability of the tunneling particle to be at a given (x,t) coordinate, our results may also be used for developing the quantum science and technology with many possible applications including but are not limited to increasing the resolution and efficiency of scanning tunneling microscopes, enhancing proton tunneling for DNA mutation and enhancing superconducting properties of junctions.

\pacs{03.65.Xp, 42.65.Sf}
\end{abstract}
\maketitle

\section{Introduction}

Quantum tunneling is studied under the branch of quantum physics whose subject is to study the processes involving atoms and photons \cite{Griffiths}. Quantum tunneling can be described as the process in which atomic particles surmount a barrier that they classically can not surmount. In classical physics, if the macroscopic objects do not have enough energy to pass over a barrier, they can not. However, in the quantum mechanical regime the atomic particles exhibit wave-like properties. Thus there is a probability that atomic particle can penetrate through the barrier. Probability of transmittance and reflectance may depend on the barrier potential width, the sign of barrier potential and the incident wavefunction. Quantum tunneling phenomena is the basis for many modern devices and technology such as the scanning tunneling microscope, quantum computing, the tunnel diode, proton tunneling for DNA mutation just to name a few. A more comprehensive introduction of quantum tunneling can be seen in \cite{Griffiths, Landau}.

Quantum tunneling is commonly studied in the frame of the linearized Schr\"{o}dinger equation (LSE)  \cite{Griffiths, Landau, Fowler}. Therefore many fascinating nonlinear phenonema is unknown or ignored by quantum mechanics community which may be used to advance the quantum science and technology.  In this paper we solve the full nonlinear Schr\"{o}dinger equation (NLSE) with a potential term to study the tunneling properties of sech type soliton and exponential (complex sinuosoid should be understood throughout this text) solutions of the NLSE for three different potential barriers. These barriers are the single rectangular barrier, the double rectangular barrier and the triangular barrier. We show that white-noise imposed on all these types of potential functions may or may not trigger modulation instability (MI) depending on the form of the incident wavefunction. If the incident wavefunction is a sech type soliton, then noisy potentials do not trigger MI and the wavefield do not turn into a chaotic one. This is expected since sech type localized solitons on a zero background interact elastically with other waves in the field and keep their original shape. This is also true for MI triggered by  a amplitude noise imposed on the wavefield directly, not only to the potential. However if the incident wavefunction is an exponential, then the noisy potential triggers MI and the wavefield turns into a chaotic one with many apparent peaks. Some of these peaks become rogue waves which are generally defined as waves with heights 2-2.2 times the significant wave height in the field. In the quantum mechanical terminology the 'rogue wavefunction' is a more suitable naming. We argue that peaks of such a chaotic wavefield generated by the noisy tunneling potentials are in the form of rational rogue wave solutions of the NLSE which may be analytically derived from a seeding exponential function using the Darboux transform formalism.

\section{Review and Numerical Solution of the Nonlinear Quantum Tunneling Problem}
Quantum tunneling can be studied in the frame of the NLSE which is given as
\begin{equation}
i\psi_t + \beta \psi_{xx} +  \zeta \left|\psi \right|^2 \psi + (V(x)-E(x))\psi =0
\label{eq01}
\end{equation}
where $\psi$ is the complex wavefunction, $x, t$ are the spatial and temporal variables and $i$ denotes the imaginary number. Typically $\beta$ is given as $\beta=-\overline{h}^2 / 2m$, where $\overline{h}$ is the reduced Planck constant and $m$ is the particle mass. Without loss of generality we set $\beta=1/2$. $V(x)$ is the potential energy of the particle, $E(x)$ is the energy of the particle associated with its motion along $x$ axis \cite{Griffiths}. For the sake of brevity we define $V(x)-E(x)=M(x)$. The sign of $M(x)$ determines the behavior of wavefunction inside the barrier. In the linear regime a negative $M(x)$ refers to evanescent behavior inside the barrier whereas a positive $M(x)$ refers to oscillatory behavior.

Quantum tunneling is generally studied in the frame of the linearized Schr\"{o}dinger equation which can be obtained by setting $\zeta=0$ in Eq.(\ref{eq01})  \cite{Griffiths, Landau, Fowler}. However this linearization causes the ignorance of many fascinating nonlinear phenomena which may be observed in nature and may possibly be used to advance quantum science and technology. In this paper we study the nonlinear quantum tunneling thus we take $\zeta=1$ and numerically solve the full NLSE using a split-step Fourier method (SSFM), which is one of the most widely used forms of the spectral methods. Like other spectral methods, in SSFM the spatial derivatives are calculated using spectral techniques. Some applications of the spectral techniques can be seen in \cite{bay2009, bay_cssfm, bayindir2016, bay_sun1, bay_sun3, bay_sun2, demiray, Karjadi2010, Karjadi2012, bayindir2016PRE, bay_nls, bayindir2016nature, bayindir2016earlyCS, bayindir2016KEE} and their broader discussion can be seen in \cite{canuto, trefethen}.  The temporal derivatives in the governing equations can be handled by different time integrating schemes such as Adams-Bashforth and Runge-Kutta etc. \cite{canuto, trefethen, demiray}, however SSFM utilizes an exponential time stepping function for the evaluation of the temporal derivatives. SSFM is based on the idea of splitting the equation into two parts, the nonlinear and the linear part and then performing the time stepping starting from the initial conditions. As a possible splitting we take first part of the NLSE as
\begin{equation}
i\psi_t= -\left| \psi \right|^2\psi -M(x) \psi
\label{eq13}
\end{equation}
which can be exactly solved as
\begin{equation}
\tilde{\psi}(x,t_0+\Delta t)=e^{i \left( \left| \psi(x,t_0)\right|^2+M(x) \right) \Delta t}\ \psi_0
\label{eq14}
\end{equation}
where $\Delta t$ is the time step and $\psi_0=\psi(x,t_0)$. The remaining part of the NLSE is
\begin{equation}
i\psi_t=-\beta \psi_{xx}
\label{eq15}
\end{equation}
Using the Fourier series approximation we obtain
 \begin{equation}
\psi(x,t_0+\Delta t)=F^{-1} \left[e^{-i \beta k^2\Delta t}F[\tilde{\psi}(x,t_0+\Delta t) ] \right]
\label{eq16}
\end{equation}
where $k$ is the Fourier transform parameter. Combining Eq.(\ref{eq14}) and Eq.(\ref{eq16}), the complete form of the split-step Fourier scheme for the NLSE with potential can be written as
 \begin{equation}
\psi(x,t_0+\Delta t)=F^{-1} \left[e^{-i \beta k^2\Delta t}F[ e^{i\left( \left| \psi(x,t_0)\right|^2+M(x) \right) \Delta t}\ \psi_0 ] \right]
\label{eq17}
\end{equation}
Starting from the chaotic initial conditions described below, the numerical solution of the NLSE with potential is obtained for later times by the SSFM scheme. This form of the SSFM scheme requires two fast Fourier transform (FFT) operations per time stepping. The number of spectral components are taken as $N=4096$ in order to make use of the FFTs efficiently. The time step is selected as $\Delta t=10^{-3}$, which does not cause any apparent stability problem. 

\section{Results and Discussion}

The processes governed by the NLSE with random initial conditions can become quite complicated. However they are still governed by a partial differential equation. Therefore for a given initial condition they can be predicted. Compared to the completely unpredictable true stochastic processes, the processes described in this study in the frame of the NLSE can be described as 'chaotic' following \cite{Akhmediev2009b, Akhmediev2011,  Akhmediev2009a, Akhmediev2014}. Although this terminology is used in different setting in the literature, throughout this paper we use the word 'chaotic' in this setting.

When the potential term is ignored, the NLSE given in Eq.(\ref{eq01}) reduces to the cubic NLSE
\begin{equation}
i{\psi}_t + \beta {\psi}_{xx} + \zeta \left|{\psi} \right|^2 {\psi} =0
\label{eq18}
\end{equation}
 Rescaling the parameters as $t= \zeta t$ and $x= \sqrt{\zeta / 2 \beta} x$ the cubic NLSE can be written in the dimensionless form
\begin{equation}
i{\psi}_t + \frac{1}{2} {\psi}_{xx} +  \left|{\psi} \right|^2 {\psi} =0
\label{eq19}
\end{equation}
It is known that the NLSE given in Eq.(\ref{eq19}) admits a soliton solution in the form of 
\begin{equation}
\begin{split}
{\psi}_0(x,t)=2A & \exp{\{-i[2x+2\left(1-A^2\right)t+\pi/2] \} }  \\
& .sech(2Ax+4At)
\label{eq20}
\end{split}
\end{equation}
where $A$ is a constant which is the amplitude of the soliton. This is one of the solutions of the NLSE whose tunneling behavior under the effect of noise are discussed in this study. The other solution of the NLSE which is used as the incident wave is the  sinusoid in the form of
\begin{equation}
\left|\psi_0 \right|=\left|\exp[i(kx)] \right|=1
\label{eq21}
\end{equation}
where $k$ denotes the initial seed plane-wavenumber which is selected as $k=1$ for all relevant incident exponential wavefunctions simulated in this paper. In the coming sections we show that white-noised quantum tunneling potentials do not trigger MI when a sech type starter is used, however they trigger MI when an exponential starter is used, thus a chaotic wavefield with many apparent is generated. This is also true for amplitude MI, amplitude modulations do not turn a wavefield with localized solution solutions into a chaotic one. However they turn the wavefield with sinusoids into a chaotic wavefield with many peaks. This occurs because envelope of the initially sinusoidal wavefield serves as a finite background for the development of rogue waves which can be obtained employing a Darboux formalism. Similar behavior can also be observed for other nonlinear models such as the Korteweg-deVries equation.

\subsection{Single Rectangular Barrier}

We start wavefield simulations using the analytical solutions of the NLSE mentioned above and with a chaotic perturbation in the form of white-noise imposed on the single rectangular barrier potential. For this purpose we define $V(x)-E(x)=M(x)$ as
\begin{equation}
M(x,t)=4 [H(x-2)-H(x+2)]+\alpha \ r_1(x,t)
\label{eq22}
\end{equation}
where $H$ is the Heaviside step function and  $i$ is the imaginary number.  $r_1$ is a uniformly distributed random number vector having values in the interval of [-1,1] \cite{Akhmediev2009a}. $\alpha$ is a parameter which shows the magnitude of noise and selected as $\alpha=1$ in this study. It is possible to add perturbations to the tunneling potential with a length scale of $L_{pert}$ through multiplying the second term in the Eq. (\ref{eq22}) by a factor of $\exp(i 2 \pi x / L_{pert})$. Or it is possible add noise perturbations with different length scales using Fourier analysis. However such a scale is not considered in the present study for illustrative purposes. The actual wavefunction is the complex $\psi$, however we only present $\left|\psi\right|$ in our simulations. 

\begin{figure}[ht!]
\begin{center}
   \includegraphics[width=3.8in]{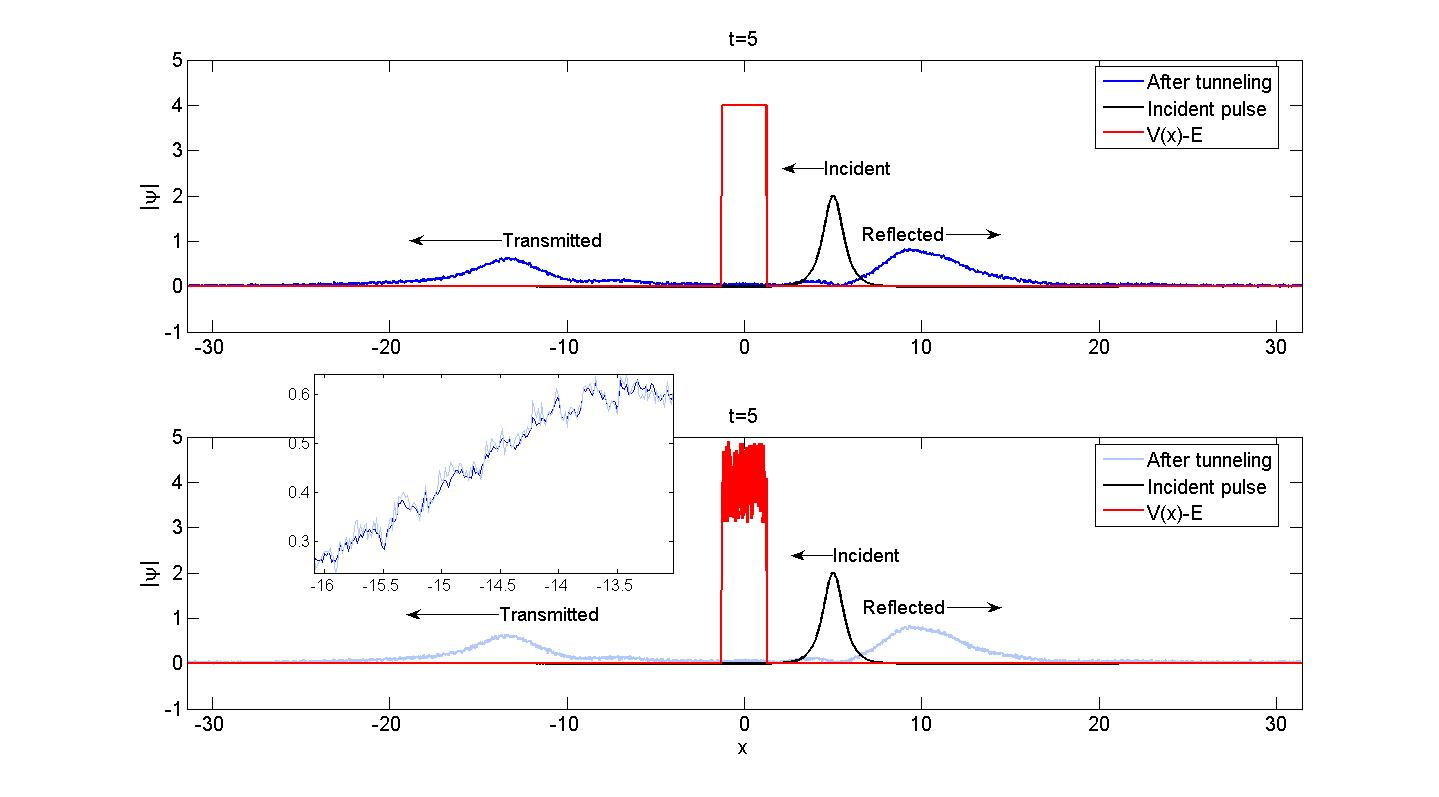}
  \end{center}
\caption{\small Quantum tunneling of $sech$ type of soliton under presence of a) single barrier potential b) white-noised single barrier potential. Note that a chaotic wavefield is not generated.}
  \label{fig1}
\end{figure}

In Fig.~\ref{fig1}, a snapshot of the numerical simulation is presented for the single rectangular barrier case for a incident wavefunction of sech type. In the first subplot, the tunneling of sech type of soliton under noise free potential is presented. The same simulation is repeated with white-noise added potential and corresponding results are depicted in the second subplot. 
A comparison of these results indicate that the noise in the quantum tunneling potential do not trigger modulation instability when a sech type soliton solution of the NLSE is used as the incident wave. This may be expected because the localized solitons on a zero background interact elastically with the wavefield, thus preserve their identity after interactions with other waves in the field. There is only a small difference in the noise levels and the noisy tunneling potential results in more amplitude noise, however well defined sech type wavefunction shape is preserved. Change in the wavelength of the transmitted and reflected waves can be observed as well, which is analogous to change in wavelength when light passes through a prism.

\begin{figure}[htb!]
\begin{center}
   \includegraphics[width=3.8in]{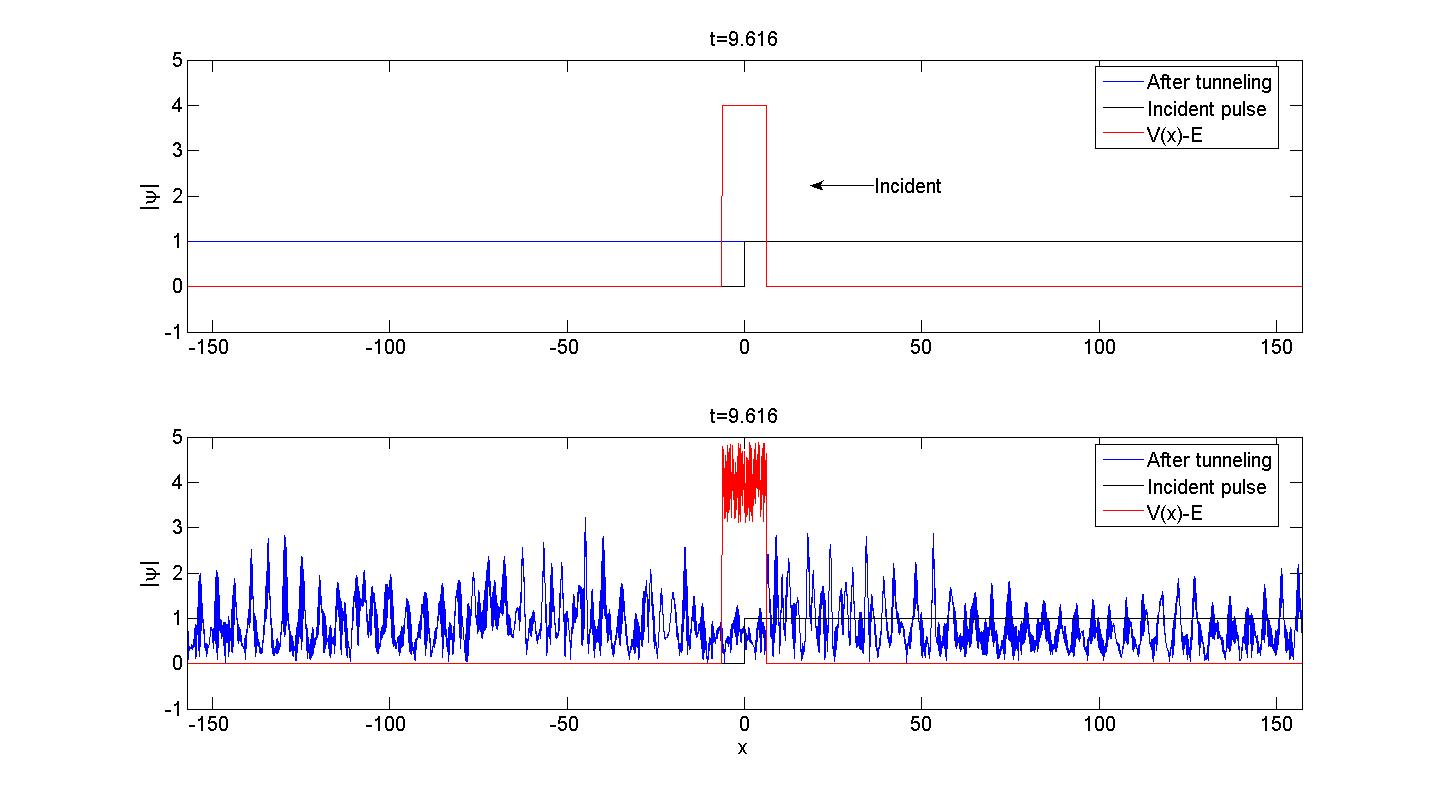}
  \end{center}
\caption{\small Quantum tunneling of exponential wavefunction under presence of a) single barrier potential b) white-noised single barrier potential. Note that a chaotic wavefield is generated.}
  \label{fig2}
\end{figure}

The snapshot of the numerical simulation shown in Fig.~\ref{fig2} is for the single rectangular barrier case for a incident wavefunction of exponential type. In the first subplot, the tunneling of exponential type of soliton under noise free potential is presented. The same tunneling problem is solved with noisy potential and corresponding results are shown in the second subplot. 
A comparison of these results indicate that noisy quantum tunneling potential trigger modulation instability when a exponential type soliton solution of the NLSE is used as the incident wave. As it will become more clear in the section, where quantum tunneling under triangular potential is discussed, that noisy potentials distort the phases of the wavefunction tunneling through barrier thus trigger a modulation instability which grows rapidly. Such a wavefield turns into a chaotic one with many peaks. Some of the peaks exceed the significant waveheight in the field with a factor of 2-2.2, thus they may be named as rogue waves, or rogue wavefunctions in the quantum mechanics terminology. The major advantage of these rogue wavefunctions is that, the tunneled particle can be located with a higher probability in the desired patches. This result may open up the way for the development or enhancing many technologies; increasing the efficiency and accuracy of the scanning tunneling microscopy, enhancing proton tunneling for DNA mutation, enhancing superconducting properties of junctions are just to name a few. Some properties of similar chaotic fields are discussed in \cite{Akhmediev2009b, Akhmediev2011, Akhmediev2009a}, however these studies mainly focuses in optics and hydrodynamics, not in the quantum mechanics branch.

\subsection{Double Rectangular Barriers}
\begin{figure}[htb!]
\begin{center}
   \includegraphics[width=3.8in]{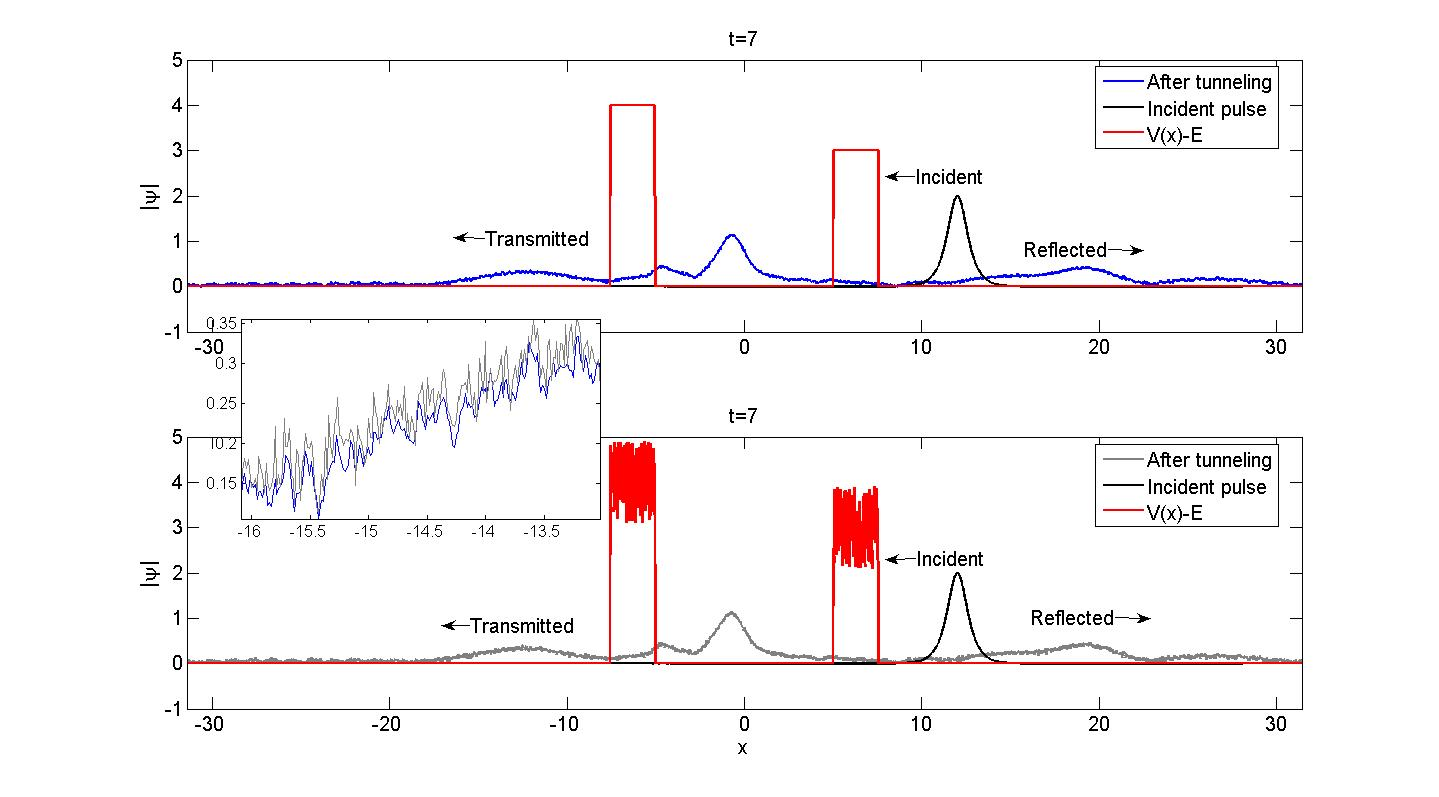}
  \end{center}
\caption{\small Quantum tunneling of $sech$ type of soliton under presence of a) double barrier potential b) white-noised double barrier potential. Note that a chaotic wavefield is not generated.}
  \label{fig3}
\end{figure}

Similar analysis can easily be extended to study the effects of multiple noisy potentials on the quantum tunneling. To show a typical configuration we consider double rectangular barrier potential which we define as
\begin{equation}
\begin{split}
M(x,t)=3 & [H(x-8)-H(x-6)] \\
& + 4[H(x+8)-H(x+6)] +\alpha \ r_1(x,t)
\label{eq23}
\end{split}
\end{equation}
where the parameter are the same with the single barrier potential case. In Fig.~\ref{fig3}, we depict a snapshot of the numerical simulation for the double rectangular barrier case. In the first subplot of this figure, the tunneling of sech type of soliton under noise free potential is presented. The same simulation is repeated with noisy potential and corresponding results are depicted in the second subplot. 

Checking Fig.~\ref{fig3}, one can realize that tunneling properties of the noisy double rectangular barrier potentials is quite similar to its single barrier analog. A comparison of these results indicate that noisy quantum tunneling potential do not trigger modulation instability when a sech type soliton solution of the NLSE is used, due to the same reasoning described above for the single barrier case. There is only a small difference in the noise levels and noisy double rectangular tunneling potential results in slightly more amplitude noise compared to the single barrier potential case, however well defined sech type wavefunction shape is again preserved. Additionally some part of the wavefunction is confined between the double barriers and it is partially transmitted and partially reflected when it interacts with the barriers. Change in the wavelength of the transmitted and reflected wavefunctions can be observed, again which is analogous to change in wavelength when light passes through multiple prisms. 

Similar to the noisy single barrier potential case, the noisy double barrier potential triggers modulation instability when a exponential type soliton solution of the NLSE is used as the incident wave. Again, some of the peaks in the chaotic wavefield exceed the significant waveheight in the field with a factor of 2-2.2, thus they may be classified as rogue wavefunctions in the quantum mechanics terminology. The major advantage of these rogue wavefunctions is that, the tunneled particle can be located with a higher probability in the desired patches of the wavefield. The striking advantages of this fact is obvious as discussed above. It is natural to expect that for a higher number of rectangular barriers the quantum tunneling behavior of noisy potentials would be similar. Using series of barrier potentials would give some flexibility to perturb certain regions of the wavefield and excite rogue wavefunctions in the desired spots or patches and avoid them in the undesired locations by manipulating tunneling potential and incident wavefunction.

\begin{figure}[htb!]
\begin{center}
   \includegraphics[width=3.8in]{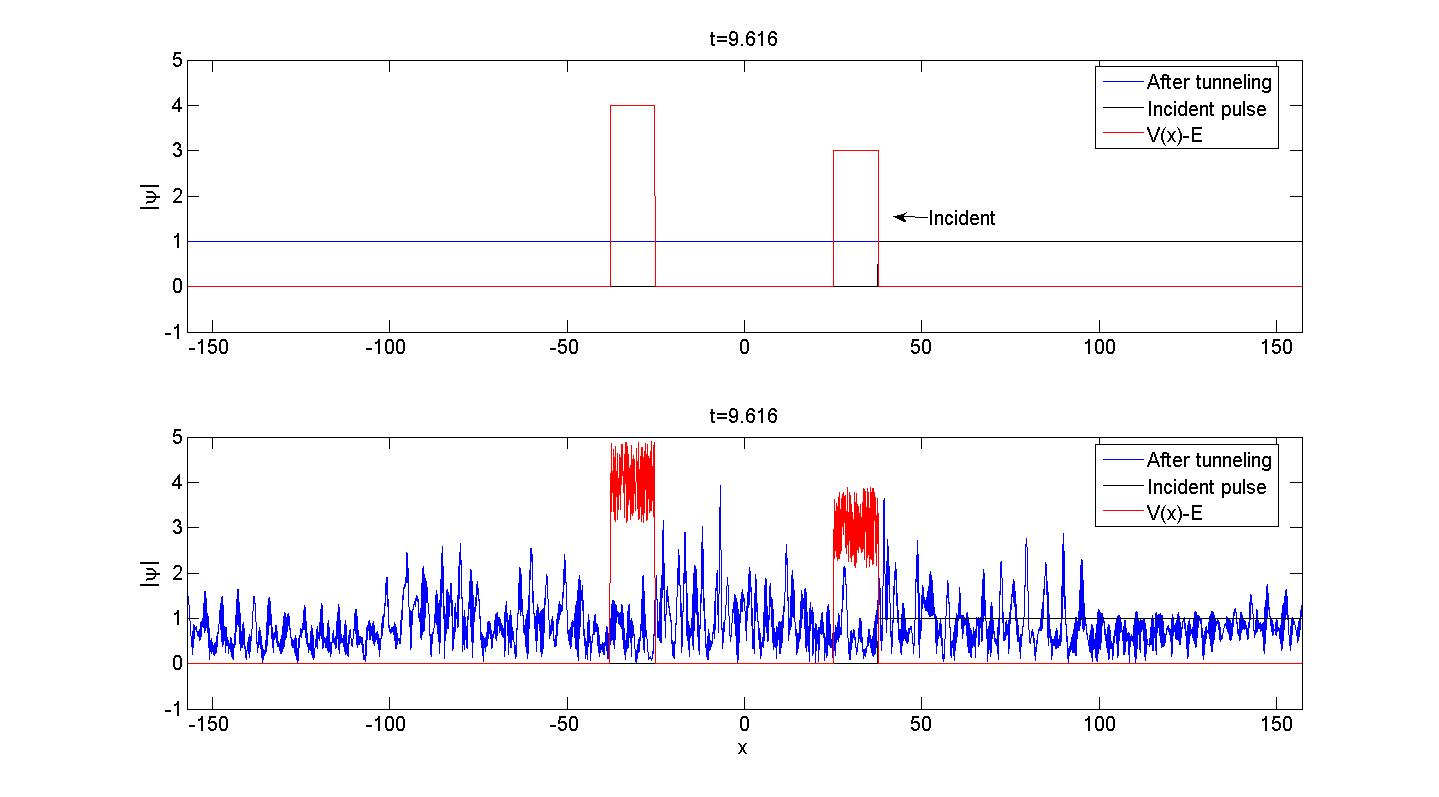}
  \end{center}
\caption{\small Quantum tunneling of exponential wavefunction under presence of a) double barrier potential b) white-noised double barrier potential. Note that a chaotic wavefield is generated.}
  \label{fig4}
\end{figure}

\subsection{Single Triangular Barrier}
Next we turn our attention to single triangular barrier potential case. In the case of triangular barrier potential, the function $V(x)-E(x)=M(x)$ can be taken as $M(x)=ax$ inside the barrier, thus the NLSE becomes
\begin{equation}
i\psi_t + \beta \psi_{xx} +  \zeta \left|\psi \right|^2 \psi + a x \psi =0
\label{eq24}
\end{equation}
where $a$ is a constant which shows the slope of the triangular barrier. It is known that a transformation of the form 
\begin{equation}
\psi(x,t)= \widetilde{\psi} \left (x- \beta a t^2 , t \right) \exp{\left( -i \left[-a xt+ \beta a^2 \frac{t^3}{3}\right] \right)}
\label{eq25}
\end{equation}
transforms the NLSE given in  Eq.(\ref{eq24}) to the standard cubic NLSE (\cite{PeregrineSmith}) given as
\begin{equation}
i\widetilde{\psi}_t + \beta \widetilde{\psi}_{xx} + \zeta \left|\widetilde{\psi} \right|^2 \widetilde{\psi} =0
\label{eq26}
\end{equation}
for which the rogue wave solutions becomes obvious \cite{Akhmediev2009b, Akhmediev2011, Akhmediev2009a}. Rescaling the parameters as $t= \zeta t$ and $x= \sqrt{\zeta / 2 \beta} x$ the cubic NLSE can be written in the dimensionless form, as given in Eq.(\ref{eq19})
One of the most early forms of the rational soliton solution of this non-dimensional NLSE is the Peregrine soliton \cite{Akhmediev2009b}. It can be written as
\begin{equation}
\widetilde{\psi}_1=\left[1-4\frac{1+2it}{1+4x^2+4t^2}  \right] \exp{[it]}
\label{eq27}
\end{equation}
where $t$ is the time and $x$ is the space parameter. It is shown that Peregrine soliton is a first order rational soliton solution of the NLSE and the second order rational soliton solution is given as \cite{Akhmediev2009b}
\begin{equation}
\widetilde{\psi}_2=\left[1+\frac{G_2+it H_2}{D_2}  \right] \exp{[it]}
\label{eq28}
\end{equation}
where
\begin{equation}
G_2=\frac{3}{8}-3x^2-2x^4-9t^2-10t^4-12x^2t^2
\label{eq29}
\end{equation}
\begin{equation}
H_2=\frac{15}{4}+6x^2-4x^4-2t^2-4t^4-8x^2t^2
\label{eq30}
\end{equation}
and
\begin{equation}
\begin{split}
D_2=\frac{1}{8} [ \frac{3}{4} &+9x^2+4x^4+\frac{16}{3}x^6+33t^2+36t^4 \\
&+\frac{16}{3}t^6-24x^2t^2+16x^4t^2+16x^2t^4 ]
\label{eq074}
\end{split}
\end{equation}
We use these analytical rogue waves to analyze the shape of the rogue wavefunctions which emerge in a chaotic wavefield. For the formulation of the noisy single triangular potential we use
\begin{equation}
M(x,t)=(1/2x+2.5)[H(x-5)-H(x+5)] +\alpha \ r_1(x,t)
\label{eq31}
\end{equation}
where the parameters are defined similarly to the rectangular barrier cases mentioned above. For the triangular barrier as the transformation given by Eq.(\ref{eq25}) confirms, the potential term of the NLSE acts like a phase shift term to the solutions of the standard cubic NLSE. Giving noise to the potentials distorts the phases of the components in the wavefield started with an incident exponential and small amplitude perturbations begin to develop. Then this wavefield turns into a chaotic one with many apparent peaks with amplitudes ($\left|\widetilde{\psi} \right|$) in the interval of  $[0-5]$, similar to the amplitude MI case as discussed in \cite{Akhmediev2009b}. Some phase patterns and properties of rogue waves are studied in the literature however the reported analysis is on purely analytical rogue wave solutions \cite{AkhmedievPhase} and do not include the chaotic wavefields. Noise is generally imposed on the amplitude of the wavefield \cite{Akhmediev2009b}, not the phases.

\begin{figure}[htb!]
\begin{center}
   \includegraphics[width=3.8in]{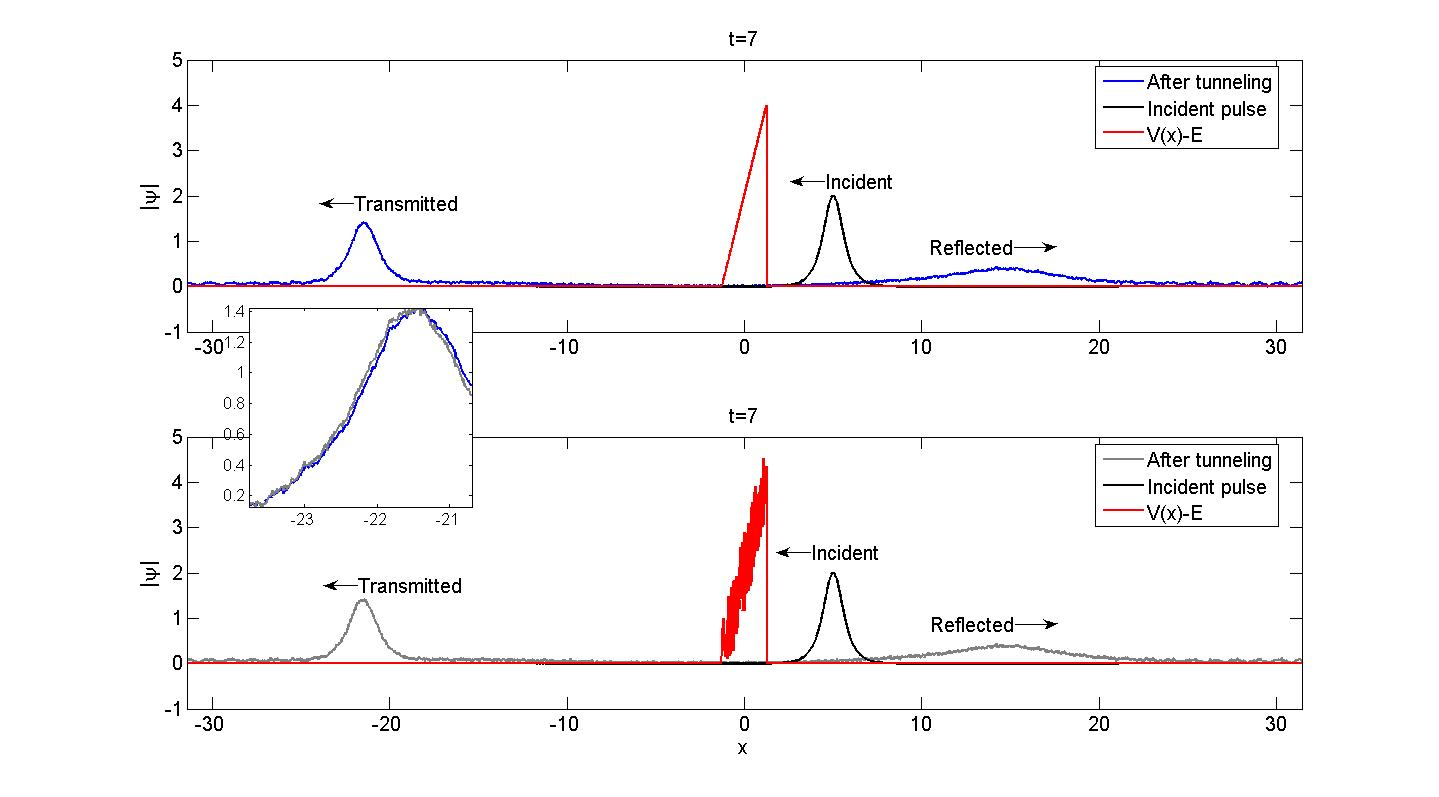}
  \end{center}
\caption{\small Quantum tunneling of $sech$ type of soliton under presence of a) single triangular potential b) white-noised single triangular potential. Note that a chaotic wavefield is not generated.}
  \label{fig5}
\end{figure}

\begin{figure}[htb!]
\begin{center}
   \includegraphics[width=3.8in]{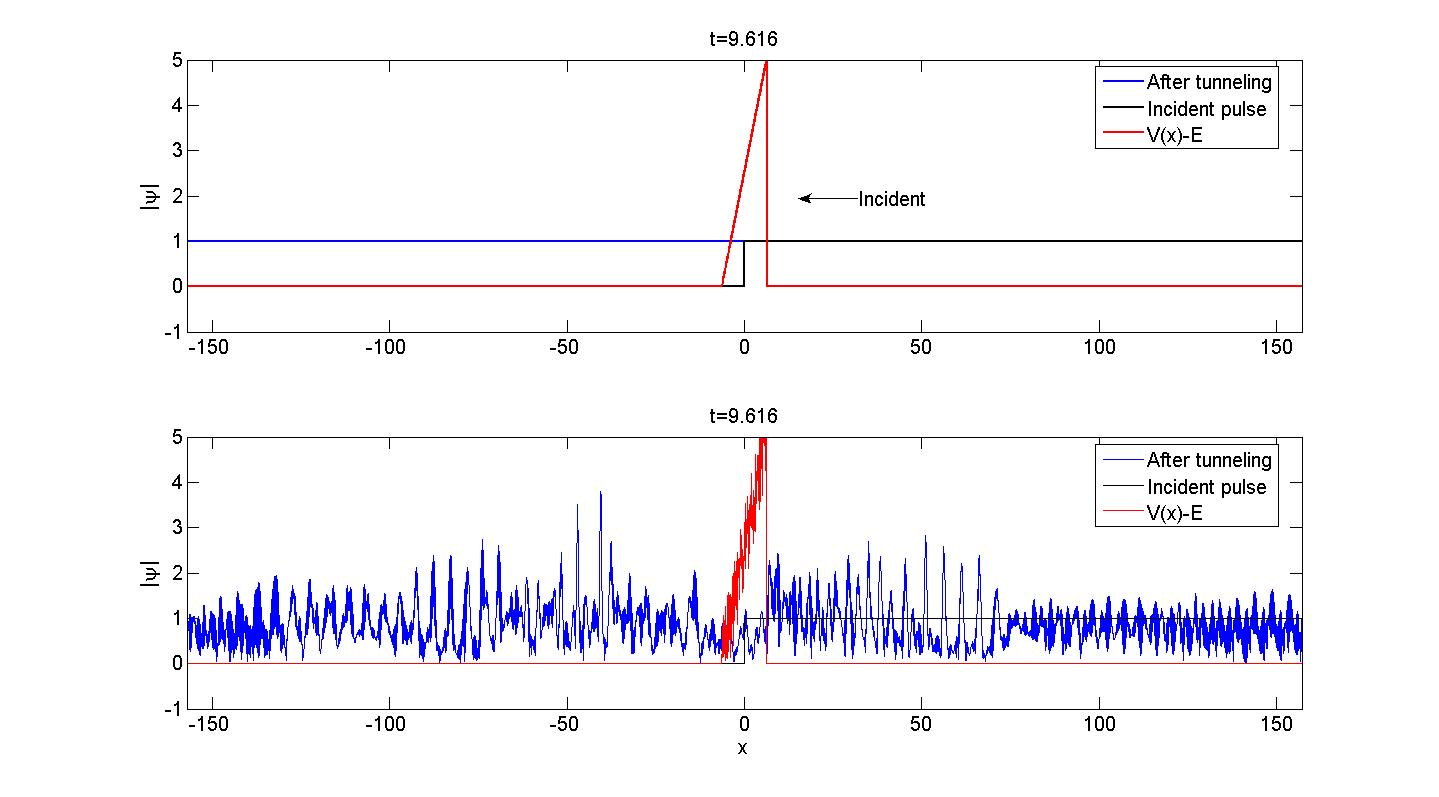}
  \end{center}
\caption{\small Quantum tunneling of exponential wavefunction under presence of a) single triangular potential b) white-noised single triangular potential. Note that a chaotic wavefield is generated.}
  \label{fig6}
\end{figure}

Checking Fig.~\ref{fig5} and Fig.~\ref{fig6} one can realize that results for the single triangular barrier potential case is quite similar to rectangular barrier potential cases. That is, a sech type incident wavefunction preserves its identity under the effect of the noisy quantum tunneling potential, however an exponential type of incident wavefunction turns into a chaotic wavefield with many peaks, some of them becoming rogue waves.

\begin{figure}[htb!]
\begin{center}
   \includegraphics[width=3.6in]{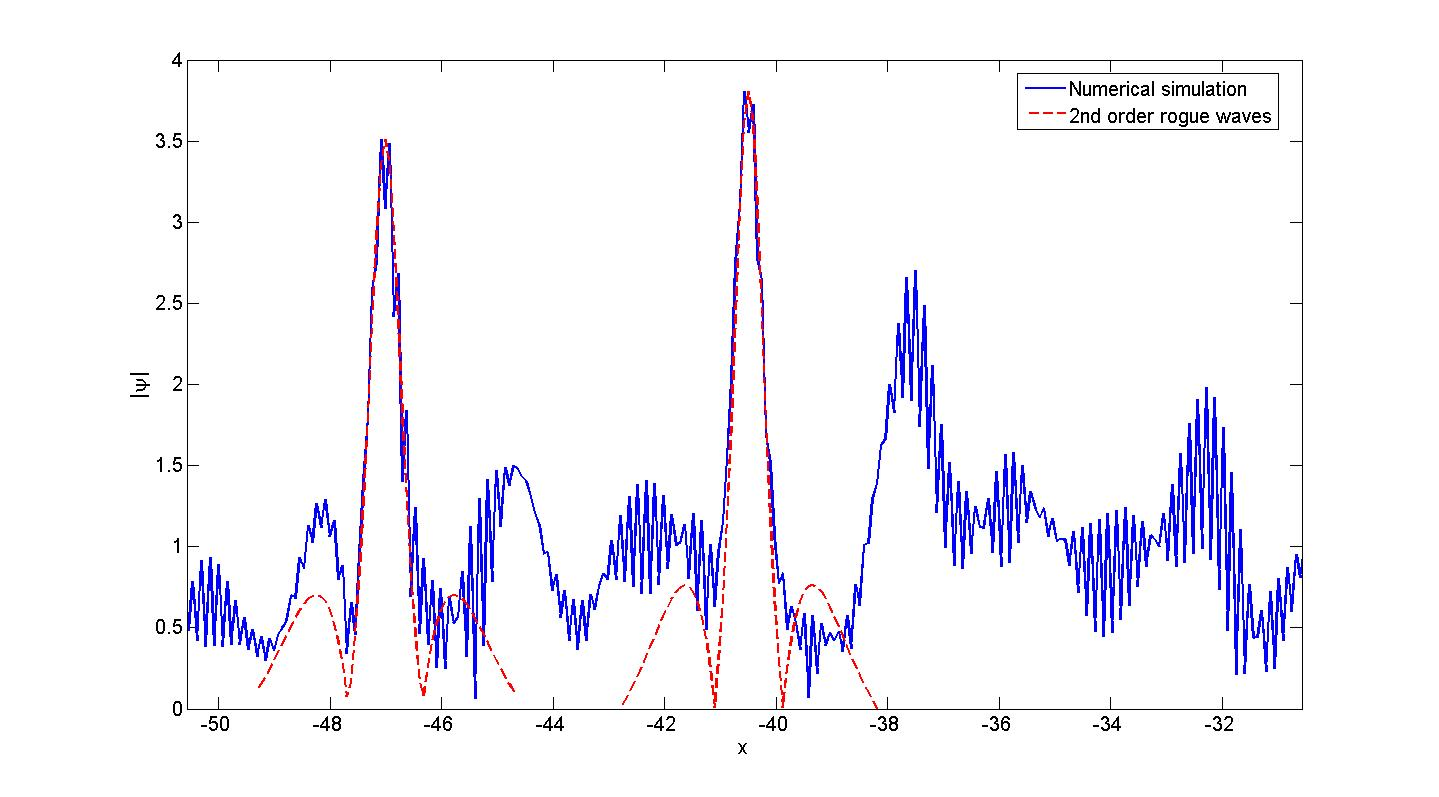}
  \end{center}
\caption{\small Comparison of the rogue wavefunctions in the chaotic field with the second order rational soliton solution of the NLSE.}
  \label{fig7}
\end{figure}

In order to analyze the shape of the rogue wave generated numerically we select the highest waves of simulations. The peak value for the apparent rogue wave in this simulation is 3.80. There is another peak in the vicinity of this highest wave and its amplitude is 3.50 (see Fig.~\ref{fig6} and Fig.~\ref{fig7}). The shape of these rogue waves can be analyzed by using the second order rational soliton solution of the NLSE \cite{Akhmediev2009a} thus we make the comparisons accordingly. In order to analyze its shape, in the Fig.~\ref{fig7} we present a comparison of the rogue wave obtained in numerical simulations with the second order rational soliton defined by Eq.(\ref{eq28}). We scale this exact second order rogue wave solution by the scaling law of the NLSE given as
\begin{equation}
\widetilde{\psi}(x,t) \rightarrow B \widetilde{\psi}(Bx,B^2t), \ \ \ B \in \Re^+ 
\label{eq32}
\end{equation}
and present the comparion in Fig.~\ref{fig7}. In this plot, the continuous blue line is taken from the numerical simulations. The exact second order rational soliton is shown by the dashed red line. The peak value for the second order rational rogue wave is 5.00 therefore we use a scaling factor of $B=3.80/5.00=0.76$ for the highest wave and $B=3.50/5.00=0.70$ for the second highest wave in the scaling law defined by Eq. (\ref{eq32}). As can be seen in Fig.~\ref{fig7}, the central part of the peaks of the numerical simulation accurately follows the exact profile. The discrepancy in the tails of the peaks are due to random smaller amplitude waves of the chaotic field that surround the peak. It is possible to conclude that the amplitude profile around the peaks closely follows the second-order rational solution. Similar results are also present in the literature for the rogue waves of the NLSE \cite{Akhmediev2009a}. Additionally it is useful to note that the rogue waves in the form the second order rational soliton can also be described by the collision of Akhmediev breathers \cite{Akhmediev2009a}, thus the rogue wavefunctions of the NLSE generated by potential noise can be described by those collusions as well. Although we analyze the shapes of the rogue wavefunctions excited by the noisy single triangular barrier potential, the results would be similar for wavefunctions excited by the noisy single or double rectangular barriers, since as a universal feature rogue waves may develop due to MI where the finite background serves as the infinite energy source which supports rogue wave formation.

The problem of nonlinear quantum tunneling under triangular potential function is identical with the problem of blocking of waves due to opposing current in the frame of the NLSE. An extended NLSE is derived by R. Smith \cite{Smith1976} to model such a nonlinear wave-current field. Recent studies on the extended NLSE of Smith shows that the probability distribution functions (PDFs) of amplitudes generated in a chaotic wave-current field are in Rayleigh distribution form \cite{bayindir2016KEE, bayindir2016ssoc} for the low amplitudes, however there is some discrepancy in the high wave amplitudes as expected which may be described by some PDFs that account for nonlinearity, such as the Tayfun distribution. Increasing the amplitude of noise, $\alpha$, do not change the PDFs since it is obvious from the transformation given by Eq.(\ref{eq25}), that the triangular potential function behaves like a phase shift term, not an amplitude modulation.

In a typical application, the rogue wavefunctions would be desired to be produced deterministically. For this purpose a time reversal generation approach (see i.e. \cite{Chabchoub}) for the development of those rogue wavefunction by imposing noise to the potential barriers can be adopted. For this purpose bias tunneling voltage can be manipulated with noise for use in quantum science or on a macroscopic scale some corrugations can be imposed in the potential barrier. It is also possible to confine the chaotic wavefield into a certain range by using multiple potentials, thus rogue wavefunctions may be excited in the desired confined regions of the wavefield. In terms of quantum mechanics terminology this means that probability of finding the tunneling particle at a given (x,t) coordinate will be higher. This property can be used to enhance many devices whose operation principles rely on quantum tunneling including but are not limited to increase the resolution and efficiency of STMs, proton tunneling for DNA mutation, enhancing superconducting properties of junctions.

\section{Conclusion}

In this paper we have studied the effects of white-noised potentials on nonlinear quantum tunneling. With this motivation, we have used a split-step scheme to numerically solve the nonlinear Schr\"{o}dinger equation with a tunneling potential. We have considered three different types of potentials. These are the single rectangular barrier, double rectangular barrier and triangular barrier potentials. For all these three cases we have showed that white-noise given to potentials do not trigger modulation instability for tunneling of the sech type soliton solutions of the NLSE. This is not only true for a wavefield excited with modulations imposed on potential, but also true for wavefields excited wşth amplitude modulation in free field conditions. However, as we have showed, the noisy potentials trigger modulation instability for tunneling of the exponential wavefunctions, thus such a wavefield turns into a chaotic one with many apparent peaks. We have argued that the peaks of such a chaotic field may be in the form of rational rogue wave solutions of the NLSE, similar to chaotic wavefields generated by the amplitude modulations. Our results can shed light upon examining the effect of noise on quantum tunneling. A rogue wavefunction means a higher probability of the tunneling particle to be at a given (x,t) coordinate, thus our results can be used for advancing the quantum science and technology. Some of the many possible applications would be increasing the resolution and efficiency of scanning tunneling microscopes, enhancing proton tunneling for DNA mutation and enhancing superconducting properties of junctions.




\end{document}